\documentclass[11pt] {article}

\usepackage{times, graphicx, bm, amssymb, amsmath, overcite}
\begin{document}


\title{Multidimensional Potentials of Mean Force from Biased Experiments Along a Single Coordinate}



\author{David D.L. Minh}


\maketitle

Department of Chemistry and Biochemistry,
Center for Theoretical Biological Physics,
and Howard Hughes Medical Institute,
University of California at San Diego, San Diego, California 92093

\begin{abstract}
External biasing forces are often applied to enhance sampling in regions of phase space which would otherwise be rarely observed.  While the typical goal of these experiments is to calculate the potential of mean force (PMF) along the biasing coordinate, here I present a method to construct PMFs in multiple dimensions and along arbitary alternative degrees of freedom. A protocol for multidimensional PMF reconstruction from nonequilibrium single-molecule pulling experiments is introduced and tested on a series of two dimensional potential surfaces with varying levels of correlation.  Reconstruction accuracy and convergence from several methods - this new protocol, equilibrium umbrella sampling, and free diffusion - are compared, and nonequilibrium pulling is found to be the most efficient.  To facilitate the use of this method, the source code for this analysis is made freely available.
\end{abstract}

\newpage



\section{\label{intro}Introduction}

External biasing forces are often applied to enhance sampling in regions of phase space 
which would otherwise be rarely observed \cite{Adcock06}.  
These perturbing forces are applied in both time-dependent, 
as in single-molecule pulling \cite{Perkins94, Florin94, Strick96, Smith96}
and steered molecular dynamics \cite{Isralewitz01, Isralewitz01a} experiments, 
or time-indepedent forms, as in umbrella sampling simulations \cite{Torrie77}.  
Usually, the direction of bias is chosen to follow the reaction coordinate 
for an interesting process, 
such as a large conformational change in a biological macromolecule.
With judicious reweighing of biased ensemble properties, 
previous workers have reconstructed the potential of mean force (PMF) along this 
bias coordinate \cite{Ferrenberg89, Roux95, Hummer01a, Hummer05, Park03, Park04, Minh06}.

The ability to reconstruct PMFs in multiple dimensions and along multiple 
alternative degrees of freedom would significantly expand the data analysis 
tool kit for biased experiments.
Although the appropriate choice of reaction coordinate is fundamental 
to kinetic calculations, transition state theory, and the accurate 
separation of distinct microstates, its selection is
limited by prior knowledge of the system.
It would be useful to compute PMFs along alternate potential reaction coordinates, 
in one or several dimensions, 
after biased data is collected along an initial, possibly arbitary choice.
In addition, by using multidimensional PMFs, the mutual entropy \cite{Cover06} 
can be used quantify and compare correlations between modes.
Here, I develop methods to perform these calculations.
They are applicable to both computer simulations, 
in which system properties are completely known, 
or laboratory experiments in which several properties can be simultaneously measured.

\section{\label{PMF}The Potential of Mean Force}

The PMF, $F_0(x)$, along a single coordinate, $x$, is defined as
\begin{eqnarray}
e^{- \beta [F_0(x) + \delta F]} =
  \frac{ \int \delta [x - x(\bm{r})] e^{ - \beta \{ H_0(\bm{r}) \} } \, d\bm{r} }{ \int e^{ - \beta \{ H_0(\bm{r'}) \} } \, d\bm{r}' }
\label{eq:PMF_1D}
\end{eqnarray}
where $\delta F$ is an arbitary constant, $\beta$ is $1/k_bT$, 
and $H_0(\bm{r})$ is the unperturbed system energy 
as a function of the phase space position $\bm{r}$.

In two dimensions, $F_0(x)$ becomes $F_0(x,y)$ and 
$\delta [x - x(\bm{r})]$ is replaced by $\delta [x - x(\bm{r})] \delta [y - y(\bm{r})]$.
Generalizing to multiple dimensions, $F_0(x)$ becomes $F_0(x,y_1,y_2,...,y_n)$ and 
the delta function is replaced by $\delta [x - x(\bm{r})] \prod_{i=1}^{n} \delta [y_i - y_i(\bm{r})]$,
where $n$ is the number of dimensions.
If the PMF encompasses all system dimensions, it can simply be referred to as the potential.

\section{\label{multidim}Multidimensional PMFs from Nonequilibrium Experiments}

Multidimensional PMFs can be reconstructed from time-dependent biasing experiments 
by applying the nonequilibrium work relation \cite{Jarzynski97a, Jarzynski97} 
in a manner analogous to Hummer and Szabo \cite{Hummer01a, Hummer05}.
In a single-molecule pulling experiment with the 
Hamiltonian $H(\bm{r},t) = H_0(\bm{r}) + V[\bm{r},t]$,
where $V[\bm{r},t]$ is the time-dependent external bias, 
the biased probability density function is
\begin{eqnarray}
 \frac{ e^{- \beta \{ H_0(\bm{r}) + V[\bm{r},t] \} } }
{ \int e^{ - \beta \{ H_0(\bm{r'}) + V[\bm{r'},0] \} } \, d\bm{r}' }
= \langle \delta (\bm{r} - \bm{r}(t)) e^{ - \beta W_t } \rangle
\label{eq:phase_space_FE}
\end{eqnarray}
where $W_t$ is the accumulated work, 
$\int_0^t \frac{\partial H(\bm{r},t')}{\partial t'} \, dt'$ \cite{Hummer01a, Hummer05}.

To calculate the PMF along the pulling coordinate, Hummer and Szabo 
consider the special case that the perturbation depends only on the 
biasing coordinate and time, $V(\bm{r},t) = V[x(\bm{r}),t]$.
They then multiply equation \ref{eq:phase_space_FE} 
by $e^{\beta V(x,t)} \delta [x - x(\bm{r})]$, 
integrate over all phase space, and use the definition in 
equation \ref{eq:PMF_1D} to obtain
\begin{eqnarray}
e^{-\beta F_0(x)} =
\langle \delta (x - x(\bm{r}(t))) e^{ -\beta \left( W_t - V[x,t] \right) } \rangle
\label{eq:pullcoord_FE}
\end{eqnarray}

If we instead multiply equation \ref{eq:phase_space_FE} by a 
two-dimensional delta function, we then obtain the two-dimensional PMF.  
Similarly, multiplying by a multidimensional delta function will lead to  
a multidimensional PMF.  (Although this is possible in principle, high 
dimensionality will lead to greater difficulty with sampling.)
In the two-dimensional case, the PMF is
\begin{eqnarray}
e^{-\beta F_0(x,y)} =
\langle \delta (x - x(\bm{r}(t))) \delta (y - y(\bm{r}(t))) e^{ -\beta \left( W_t - V[x,t] \right) } \rangle
\label{eq:twoD_FE}
\end{eqnarray}

In any particular time slice, the position will be more highly sampled near 
the minima of the bias potential.
To improve accuracy over the entire range of $x$, data from different time slices 
can be combined by adapting the 
weighted histogram analysis method (WHAM) \cite{Ferrenberg89}.
After unbiasing with respect to the accumulated work as well as the time-dependent potential
\cite{Hummer01a, Hummer05}, 
the resultant PMF is 
\begin{eqnarray}
F_o(x,y) = -\beta^{-1} \ln \frac{ {\sum_{t}
\frac {\langle \delta (x - x(\bm{r}(t))) \delta (y - y(\bm{r}(t))) e^{ -\beta W_t } \rangle}
{\langle e^{ -\beta W_t } \rangle}} }
{ {\sum_{t} \frac {e^{ -\beta V(x,t) }}
{\langle e^{ -\beta W_t } \rangle}} }
\label{eq:time_slice}
\end{eqnarray}

\section{\label{Correlation}Correlation}

The usefulness of reconstructing $y$ from experiments biased along $x$ is 
expected to be related to their degree of correlation.  
One standard measure for the correlation of two coordinates with known 
probability density, $f(x,y)$, is the mutual entropy \cite{Cover06},
\begin{eqnarray}
S_{mutual} = \int \int f(x,y) \log \frac {f(x,y)}{f(x)f(y)} \, dx \, dy
\label{eq:mut_entropy}
\end{eqnarray}

where $f(x)$ and $f(y)$ are the marginal probabilities along 
$x$ and $y$.  
As a reference, let $f^*(y)$ be a probability density estimate
based on a histogram of sampling along $y$ without any correction
for external bias along $x$.
Let the estimate of the PMF using $f^*(y)$, $F(y) = -k_bT\ln(f^*(y))$, 
be known as the na\"\i ve estimate.
If $x$ and $y$ are completely independent, $f(x,y) = f(x)f(y)$,
$f(y) = f^*(y)$, and the mutual entropy is zero.
In this case, a na\"\i ve estimate of the PMF along y
should be as good as a reconstruction from equation \ref{eq:time_slice}.
If $x$ and $y$ are highly correlated, then the mutual entropy should be high.
The na\"\i ve estimate will fail, making the utility of 
equation \ref{eq:time_slice} more evident.

\begin{figure}
\includegraphics{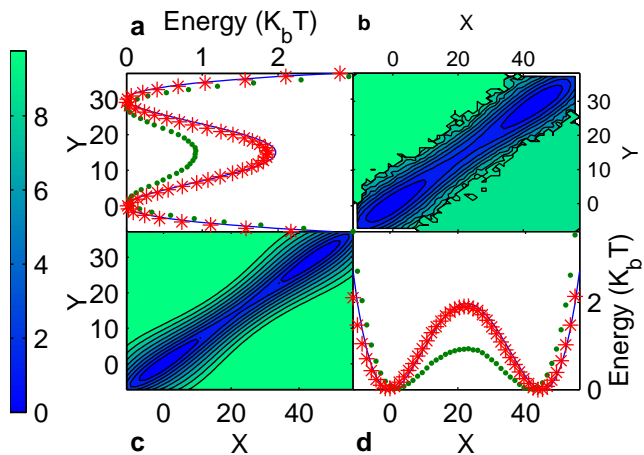}
\caption{\label{fig:u0}
PMF reconstruction on the surface with $\alpha = 0$
from umbrella sampling.
\textbf{a}. PMFs along the $y$ coordinate.  The blue line is the theoretical 
surface, the red asterisks are the biased reconstruction, and 
green dots are the na\"\i ve reconstruction.
\textbf{b}. Contour plot of the reconstructed potential.
\textbf{c}. Contour plot of the theoretical potential.  
The scale for both b and c and indicated by the side color bar.
\textbf{d}. PMFs along the $x$ coordinate.  Annotation is the same as in \textbf{a}.}
\end{figure}

\begin{figure}
\includegraphics{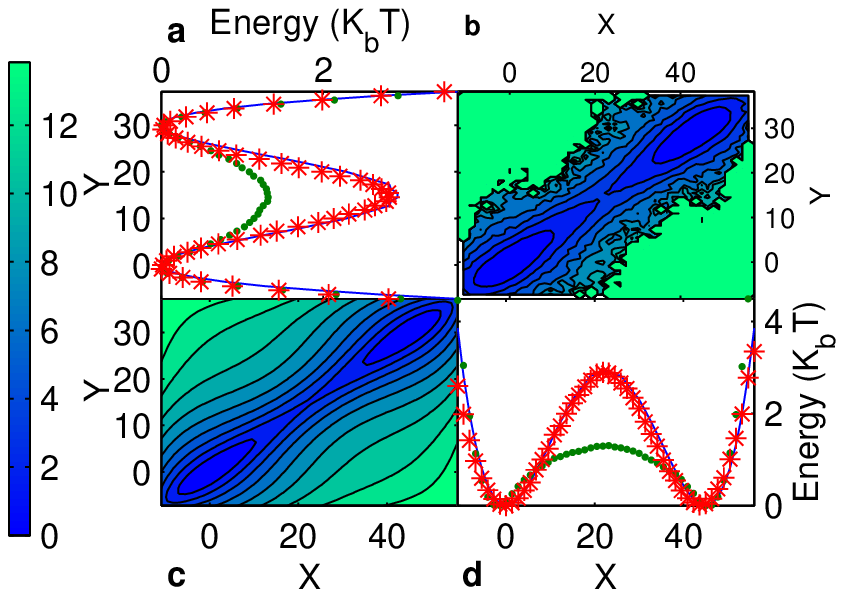}
\caption{\label{fig:p30}
PMF reconstruction on the surface with $\alpha = 0.3$ 
from time-dependent biasing experiments.
Annotation is the same as in Figure \ref{fig:u0}.}
\end{figure}

\begin{figure}
\includegraphics{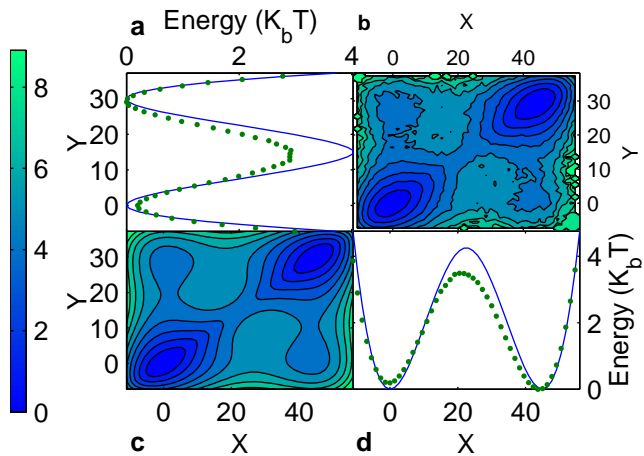}
\caption{\label{fig:d75}
PMF reconstruction on the surface with $\alpha = 0.75$
from diffusion experiments.
Annotation is the same as in Figure \ref{fig:u0}, 
except there are no WHAM reconstructions.}
\end{figure}

\begin{figure}
\includegraphics{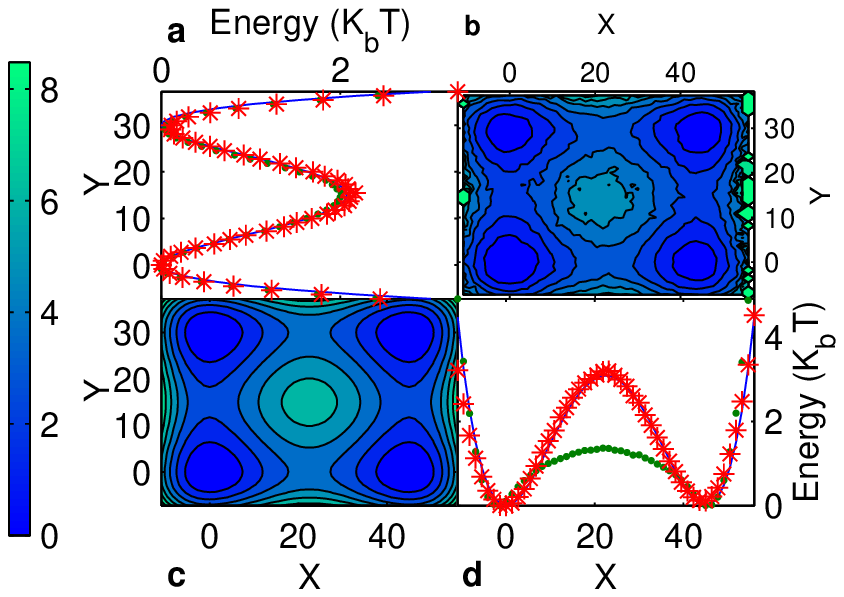}
\caption{\label{fig:p100}
PMF reconstruction on the surface with $\alpha = 1.0$
from time-dependent biasing experiments.
Annotation is the same as in Figure \ref{fig:u0}.}
\end{figure}

Equation \ref{eq:time_slice} was tested on series of 
two-dimensional potential surfaces consisting of a linear combination of
\begin{eqnarray}
U_a(x,y) = 5 x^2 (x-45)^2 + 1.7 \times 10^{-4} y^2 (y-30)^2 \label{eq:Ua}
\end{eqnarray}
and
\begin{eqnarray}
U_b(x,y) = 15 \ln[(2x - 3y)^2 + 100] + 1.25 \times 10^{-7}(2x+3y)^2(2x+3y-180)^2 \label{eq:Ub}
\end{eqnarray}
according to
\begin{eqnarray}
U(x,y,\alpha) = \alpha U_a + (1 - \alpha)U_b
\label{eq:U}
\end{eqnarray}

where $U$ is in units of $pN \cdotp nm$ and positions $x$ and $y$ are in units of $nm$.
$U_a$ and $U_b$ both have minima at (0,0) and (45,30), 
and $U_a$ has additional minima at (0,30) and (45,0).
Part c of figures \ref{fig:u0}, \ref{fig:p30}, \ref{fig:d75}, and \ref{fig:p100} show 
the potential $U(x,y,\alpha)$ at $\alpha$ of 0, 0.30, 0.75, and 1.00, respectively.

\begin{figure}
\includegraphics{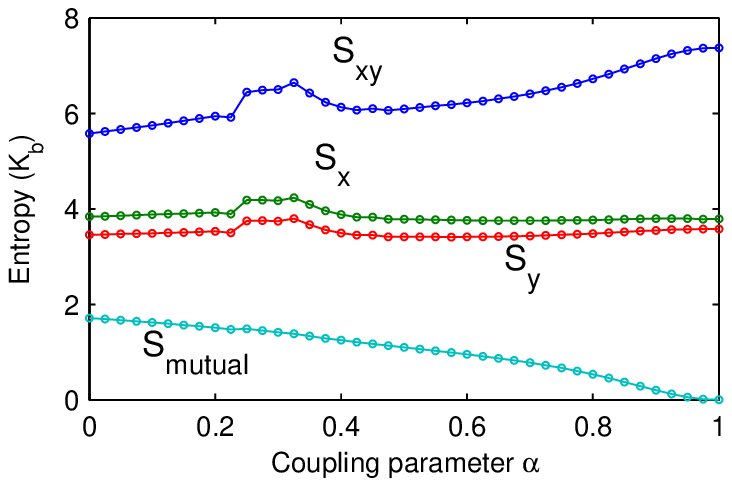}
\caption{\label{fig:mutentropy}
Entropy as a function of the coupling parameter $\alpha$.  
The mutual entropy, $S_{mutual}$, is designated by connected squares.
The joint entropy, $S_{xy}$, and the marginal entropies, $S_x$ and $S_y$,
are labeled and marked by connected circles.}
\end{figure}

In $U_a$ (Fig. \ref{fig:p100}c), $x$ and $y$ are completely independent, while in 
$U_b$ (Fig. \ref{fig:u0}c), they are highly correlated.  This is quantified by the 
mutual entropy (Fig. \ref{fig:mutentropy}), which was calculated by assuming a
Boltzmann distribution for $x$ and $y$, $f(x,y) = e^{-\beta U(x,y)}$, 
and numerically integrating equation \ref{eq:mut_entropy} 
between $-20<x<65$ and $-20<y<50$ via adaptive Simpson quadrature.

As expected, mutual entropy decreases as the potential becomes more like $U_a$.
The joint entropy, $S_{xy} = \int f(x,y) \, dx \, dy$, and the 
marginal entropies, $S_{x} = \int f(x) \, dx$ and $S_{y} = \int f(y) \, dy$,
are mostly stable as a function of $\alpha$.  
These magnitude of these factors cannot be directly compared 
to the mutual entropy because they are offset by an arbitrary constant, 
whereas mutual entropy is invariant to scale.

\section{\label{Brownian Dynamics}Simulations on a Two-Dimensional Energy Surface}

Brownian dynamics simulations \cite{Ermak78} were run 
on the series of potential surfaces.
The progress of a dynamical variable is given by 
$\Delta x(t) = \frac {F(x(t-1)) D \Delta t }{ k_bT } + (2 D \Delta t)^{1/2} R$, 
where $F(x) = \frac {- \partial U(x,y)}{\partial x}$ is the force,
$T$ is the temperature (300 K),
$D$ is the diffusion constant (1200 $nm^2/s$), 
$\Delta t$ is the time step (1 ms),
and R is a normally distributed random variable.

After 1000 steps of equilibration, umbrella sampling and pulling simulations 
were run for 100 iterations of 10000 time steps.
In addition, 10 unbiased diffusion simulations were run for 100000 time steps.
Umbrella sampling was performed with a harmonic bias, 
$V(x(\bm{r})) = 1/2 k_s (b - x(\bm{r}))^2$, setting the spring constant, 
$k_s$, at $0.1 pN/nm$ and the bias center, $b$, at 
one of 100 evenly spaced positions between -5 and 50 nm.
Pulling simulations were performed with the time-dependent perturbation,
$V[x(\bm{r}),t] = 1/2 k_s (b(t) - x(\bm{r}))^2$, with $k_s$ set to $0.4 pN/nm$.
A periodic biasing program \cite{Braun04},
$b(t) = 22.5 - 27.5 \cos (2 \pi t/10000)$, 
which has one period per trajectory, was used.
Accumulated work was numerically integrated by
$W_t = \sum_{j=1}^{t}-k_s[(x_j + x_{j-1})/2 - (b_j + b_{j-1})/2](b_j - b_{j-1})$, where
$b_j$ is the position of the bias center at time step $j$.

Theoretical surfaces, $U(x)$ and $U(y)$, were calculated by 
assuming a Boltzmann distribution, $f(x,y) = e^{-\beta U(x,y)}$, and
numerically integrating $f(x,y)$ over the 
other coordinate (with limits of $-20<x<65$ and $-20<y<50$) 
using adaptive Simpson quadrature to obtain a marginal probability density.
PMFs are then obtained by taking natural logorithms and multiplying by $-k_bT$.
In time-dependent biasing experiments, reconstructed surfaces 
for $x$ were calculated using equation \ref{eq:pullcoord_FE},
two-dimensional surfaces from equation \ref{eq:time_slice},
and surfaces along $y$ by integrating over $x$; 
from the estimate of $f(x,y)$, the sum is taken over $x$ to yield an 
unnormalized $f(y)$, from which $U(y) = -k_bT \ln f(y)$ is calculated.
PMFs from umbrella sampling were calculated as described by Roux \cite{Roux95}.
Minima of all PMFs were set to zero.  
Errors were calculated by the root mean square deviation 
from the theoretical surface at 50 points 
between $-10.9091<x<55.9091$ and $7.2727<y<37.2727$, 
ranges which include the zeroes of $U_a$ and $U_b$.  
The root mean square error (RMSE) along $y$ is defined as 
$\sqrt{\frac{1}{50} \sum_{i=1}^{50} [F_{theoretical}(y_i)- F_{reconstructed}(y_i)]^2}$.

\section{\label{results}Results and Discussion}

\begin{figure}
\includegraphics{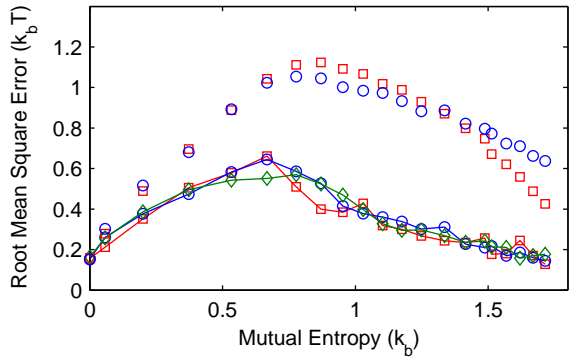}
\caption{\label{fig:error}Root mean square error for $y$ PMF estimates 
as a function of mutual entropy.
Reconstruction errors for pulling experiments are indicated by squares and
for umbrella sampling by circles.  Errors from unbiased diffusion on the surface 
are indicated by diamonds.
Errors in WHAM-reconstructed surfaces are connected by solid lines,
while the na\"\i ve estimate errors are not connected.}
\end{figure}

When $x$ and $y$ are completely independent, the na\"\i ve estimate for $F(y)$ is 
as good as the estimate from equation \ref{eq:time_slice} (Fig. \ref{fig:p100}a).
If $x$ and $y$ are highly correlated, however, the na\"\i ve estimate fails and 
a WHAM-informed reconstruction method is necessary (Fig. \ref{fig:u0}).  
There is no simple dependence, however, between error and 
mutual entropy (Fig. \ref{fig:error}), as other factors of the free energy surface 
come into play.

For example, due to the highly diffusive nature of the dynamics 
and the large timestep necessary for fast sampling, 
sampling is erroneously improved near the boundaries of small low-energy wells.
This leads to systematic free energy underestimates in all simulations, even those 
without any external bias (i.e. Fig. \ref{fig:d75}).  Although there are errors 
in WHAM-informed reconstructions from simulations with external bias, 
they are no worse than errors from long simulations with free diffusion (Fig. \ref{fig:error}).

On certain free energy surfaces, phase space sampling 
along the $y$ dimension may be limited by trapping in local minima.
This is especially true if $x$ and $y$ are relatively independent 
and biasing along $x$ does not enhance conformational sampling along $y$.
Sampling is generally limited in high-energy regions of 
phase space (see Figs. \ref{fig:u0}, \ref{fig:p30}, \ref{fig:d75}, and \ref{fig:p100}).
In these regions, the most accurate two-dimensional PMF reconstruction would
require biasing along both $x$ and $y$, which can be accounted for by a simple modification 
of equation \ref{eq:time_slice}.
In general, accurate PMF reconstruction in higher dimensions may require biasing 
along coordinates that closely correspond to these dimensions.
Close correspondence may be indicated by high mutual entropy; this 
aspect of analysis will be left to further study.

\begin{figure}
\includegraphics{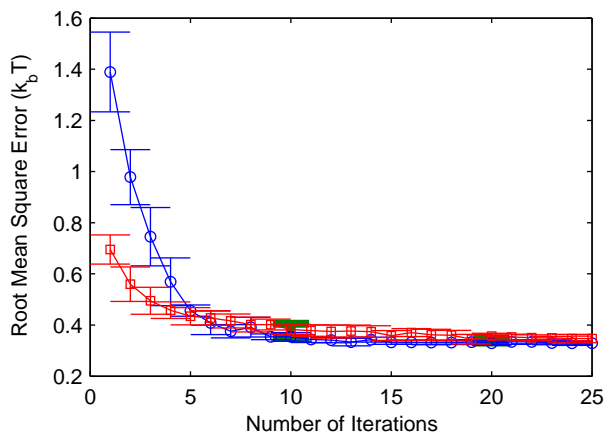}
\caption{\label{fig:convergence}
Root mean square error for $y$ PMF estimates as a function of 
number of simulations, $n$.
Error bars indicate the mean and standard deviations of RMSE 
from five reconstructions using $n$ simulations.
Reconstruction errors for pulling experiments are indicated by squares and
for umbrella sampling by circles.
Unbiased simulations were run for 10 times as long and are therefore 
spaced 10 times farther apart on the x axis.}
\end{figure}

On the toy two-dimensional surfaces studied, PMF reconstructions along $y$ using 
equation \ref{eq:time_slice} actually converge more quickly than
equilibrium umbrella sampling and reconstruction from unbiased 
experiments (Fig. \ref{fig:convergence}).
While time-dependent biasing experiments will sample the whole span of the bias coordinate
\textit{de facto}, umbrella sampling will inherently require a certain number of simulations at 
different bias centers to attain this broad range.  
Similarly, unbiased simulations will need to be run for longer in order to reach 
more distal regions of phase space.
Generally, it is preferable to run a greater number of nonequilibrium experiments than a 
single long equilibrium experiment because the outcome of each individual trajectory 
can be highly dependent on initial conditions.
These sampling problems described on two-dimensional surfaces are likely to 
be enhanced in systems with higher complexity.
The main caveat with this method, as with all methods based on the nonequilibrium work relation, 
is that with greater free energy differences, the likelihood of observing a
trajectories with negative dissipative work (work that is less than the free energy difference) 
is increasingly unlikely.

Pulling experiments are poised to find broader utility in 
studies of nanoscale systems, particularly when free energy differences between 
states are not prohibitively large.
I hope this work will find wide application in computational and laboratory experiments 
and stimulate further research in the area.
MATLAB source code to perform the described analysis 
is available at http://mccammon.ucsd.edu/\~{}dminh/software/.

\section{Acknowledgments}
I would like to thank R. Amaro, C.E. Chang, J. Gullingsrud, 
and J.A. McCammon for helpful discussions.
I am funded by the NIH Molecular Biophysics Training Grant at UCSD.
McCammon group resources are supported by the NSF (MCB-0506593), NIH (GM31749), 
HHMI, CTBP, NBCR, W.M. Keck Foundation, and Accelrys, Inc.

\end{document}